\documentclass[cameraready]{Interspeech}

\title{ZeSTA: Zero-Shot TTS Augmentation with Domain-Conditioned Training for Data-Efficient Personalized Speech Synthesis}

\author[affiliation={1}, orcid=0009-0005-2528-9263]{Youngwon}{Choi}
\author[affiliation={2}]{Jinwoo}{Oh}
\author[affiliation={1}]{Hwayeon}{Kim}
\author[affiliation={1}, orcid=0009-0007-3079-0647]{Hyeonyu}{Kim}

\address{
    $^1$ Maum AI Inc., Republic of Korea \\
    $^2$ Humelo Inc., Republic of Korea
}

\email{youngwonchoi@maum.ai, jwoh@humelo.com, khy0908@maum.ai, hykim@maum.ai}


\usepackage{comment}
\usepackage{multirow}
\usepackage{cite}
\usepackage{tikz}
\usepackage{subcaption}

\begin{document}

\maketitle

\begin{abstract}
We investigate the use of zero-shot text-to-speech (ZS-TTS) as a data augmentation source for low-resource personalized speech synthesis.
While synthetic augmentation can provide linguistically rich and phonetically diverse speech, naively mixing large amounts of synthetic speech with limited real recordings often leads to speaker similarity degradation during fine-tuning.
To address this issue, we propose ZeSTA, a simple domain-conditioned training framework that distinguishes real and synthetic speech via a lightweight domain embedding, combined with real-data oversampling to stabilize adaptation under extremely limited target data, without modifying the base architecture.
Experiments on LibriTTS and an in-house dataset with two ZS-TTS sources demonstrate that our approach improves speaker similarity over naive synthetic augmentation while preserving intelligibility and perceptual quality.
Audio samples are available on our web page\footnote{https://zeroone-universe.github.io/zesta/}.
\end{abstract}

\section{Introduction}

Recent neural text-to-speech (TTS) models have achieved near human-level naturalness under sufficient training data, including lightweight architectures suitable for practical deployment~\cite{renfastspeech,kim2021conditional,lim2022jets, kim2020glow}.
With these advances, personalized TTS, which adapts a model to a specific target speaker, has gained increasing attention with the growing demand for custom voices~\cite{chenadaspeech, min2021meta}.
However, adapting a model to previously unseen speakers remains challenging, particularly when only limited target-speaker recordings are available, motivating data-efficient adaptation methods.

Following the categorization in~\cite{hong2024leveraging}, existing approaches to personalized TTS can be broadly grouped into zero-shot and fine-tuning paradigms.
Recent zero-shot TTS (ZS-TTS) models are built upon large-scale generative modeling frameworks to generate the voices of unseen speakers without additional training, showing impressive generalization~\cite{le2023voicebox, borsos2023audiolm, kharitonov2023speak}.
However, such models are often computationally demanding for practical deployment. 
Meanwhile, lightweight ZS-TTS models based on conventional TTS acoustic modeling~\cite{casanova2021sc, casanova2022yourtts, lee2022hierspeech} can be deployed more easily but have been reported to achieve speaker similarity lower than large-scale generative ZS-TTS models~\cite{le2023voicebox}.
Adaptation via fine-tuning~\cite{chenadaspeech, mehrish2023adaptermix, huang2022meta}, on the other hand, can produce high-fidelity speech when adequate target-speaker data are available, yet its performance is highly sensitive to data scarcity.

In low-resource scenarios where target-speaker recordings are extremely limited, fine-tuning with data augmented by ZS-TTS may offer a promising solution for building lightweight personalized models suitable for practical deployment.
However, principled strategies for incorporating synthetic speech into such low-resource fine-tuning settings remain largely underexplored.
We observe that naively mixing large amounts of ZS-TTS speech with scarce target-speaker data improves intelligibility while degrading speaker similarity.
To address this challenge, we propose ZeSTA, a simple domain-conditioned training framework that distinguishes between real and synthetic speech with a small additional embedding and employs real-data oversampling to stabilize adaptation, without modifying the base TTS architecture.
Objective evaluations on LibriTTS and the in-house dataset with multiple ZS-TTS models demonstrate that the proposed approach preserves speaker similarity while retaining intelligibility gains from synthetic augmentation. 
Subjective evaluations further confirm improved perceptual speaker similarity without degrading speech naturalness.

\section{Related Works}

Previous studies have explored generating synthetic speech samples to expand training data for TTS systems.
Voice conversion (VC) methods~\cite{huybrechts2021low, comini2022low, ribeiro2022cross} have been used to create additional target-speaker utterances without parallel data.
However, these approaches typically require training or adapting a VC model using recordings from the target speaker, which may limit their practicality in low-resource personalization scenarios.
Separately, several works~\cite{sharma2020strawnet, hwang2021tts, song2022tts} have utilized TTS models for data augmentation of TTS itself, where an autoregressive TTS system generates synthesized speech to train non-autoregressive or lightweight models.
Such approaches often rely on sufficient data to train a capable teacher model and do not explicitly address scenarios with extremely limited target-speaker recordings.

Meanwhile, several studies~\cite{nespoli2023zero, bae2025generative, kim2025data} have used ZS-TTS to generate personalized or task-specific speech for downstream applications, highlighting its ability to provide diverse and controllable synthetic data.
Motivated by this potential, we explore the use of ZS-TTS as an augmentation source for personalized TTS under low-resource conditions.

\begin{figure}[t]
  \centering
  \includegraphics[width=\linewidth]{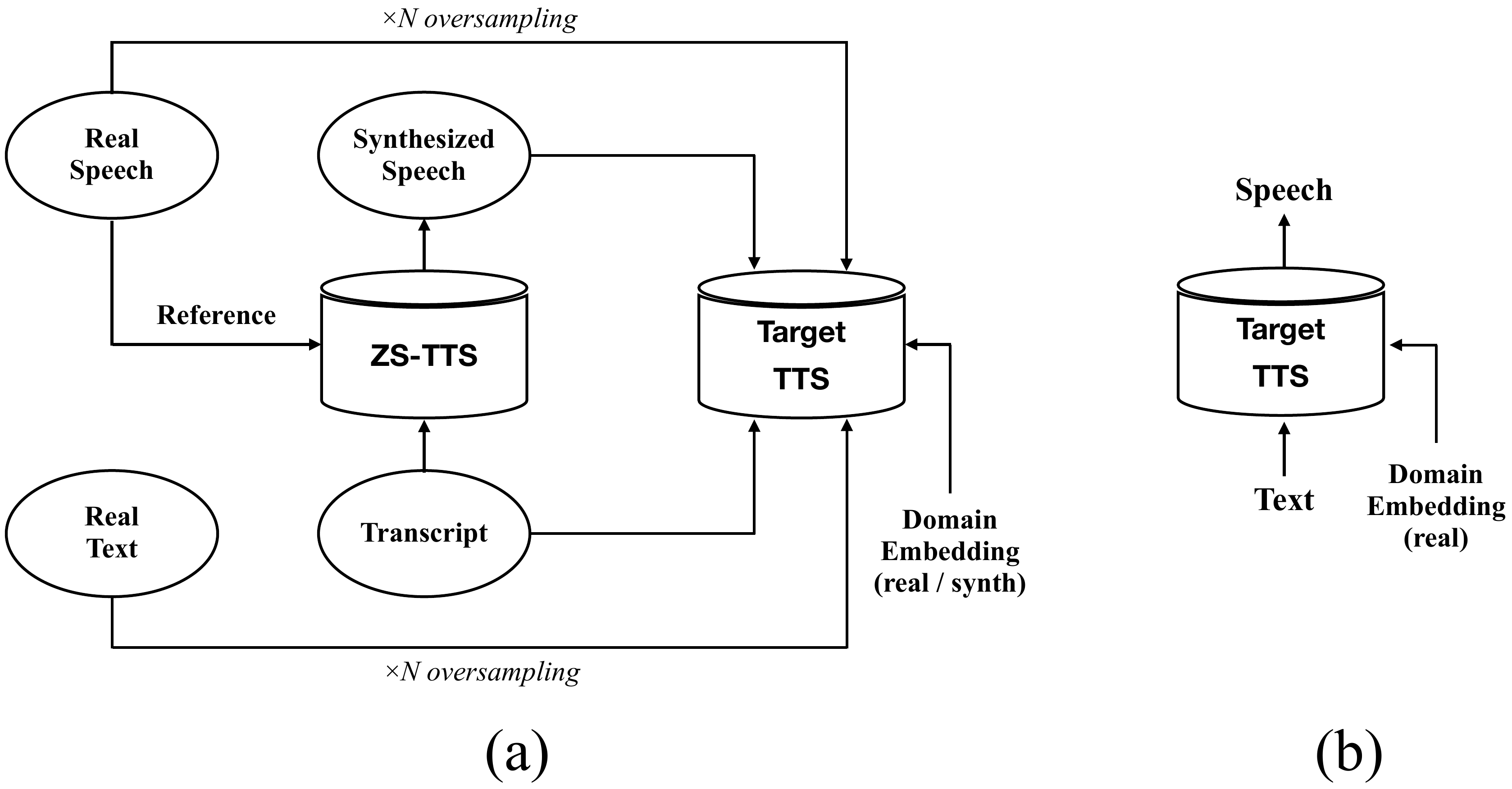}
  \caption{ZeSTA framework: training pipeline (a) and inference pipeline (b).}
  \vspace{-0.7em}
  \label{fig:fig1}
\end{figure}

\section{Method}

\subsection{Zero-Shot Speech Synthesis for Data Augmentation}
To mitigate data scarcity in personalized TTS, we leverage publicly available ZS-TTS models as external data generators to synthesize additional speech for adaptation.
Given a limited set of reference utterances from a target speaker, the source ZS-TTS models generate speech conditioned on the target speaker’s style while preserving the linguistic content of the input text.
While speech samples generated with ZS-TTS are typically highly stable and intelligible, they often exhibit reduced speaker similarity to the target speaker, as reflected by the speaker embedding cosine similarity scores in Table~\ref{tab:dataset_analysis}.
When large amounts of such synthetic speech are used during fine-tuning, this mismatch can bias the model toward synthetic-domain characteristics, highlighting the need to account for the domain discrepancy between real and synthetic speech during adaptation.

\subsection{Domain-Conditioned Training}
To address this domain discrepancy, we adopt a simple domain-conditioned training (DC) strategy that explicitly encodes the data origin of each training sample.
Under DC, TTS adaptation can be interpreted as optimizing a conditional probability $p(y \mid x, d)$, where $x$ denotes the input text, $d \in \{\text{real}, \text{synthetic}\}$ indicates the domain of each training sample, and $y$ denotes the target speech.
At inference time, synthesis is performed by conditioning on $d=\text{real}$.

From an architectural perspective, this formulation aligns with modern multi-speaker TTS systems~\cite{kim2021conditional, kim2020glow, chen2020multispeech}, where linguistic encoding primarily captures speaker-independent phonetic information, while speaker-related factors are injected into the acoustic generation module.
Specifically, the text encoder $f_{\text{text}}(\cdot)$ maps the input text $x$ to a speaker-agnostic linguistic representation $h_{\text{ling}} = f_{\text{text}}(x)$, and the acoustic generation module $g(\cdot)$ generates synthesized speech $\hat{y}$ conditioned on both $h_{\text{ling}}$ and the domain label $d$, i.e., $\hat{y} = g(h_{\text{ling}}, d)$.
As a result, linguistic augmentation effects provided by synthetic speech are retained through the text representation, while domain-specific acoustic characteristics are modulated via the domain label, effectively mitigating speaker identity drift.

\subsection{Real-Data Oversampling}
While domain-conditioned training effectively mitigates the bias introduced by abundant synthetic speech, we observe that speaker similarity can be further enhanced by modestly emphasizing real target-speaker samples during fine-tuning.
Specifically, we oversample (OS) real utterances by a small factor, which consistently improves speaker similarity without altering the model architecture or inference procedure.
Figure~\ref{fig:fig1} illustrates the training and inference pipelines of the ZeSTA framework.

\begin{table}[t]
\centering
\footnotesize
\caption{Summary statistics of the LibriTTS and YoBind datasets. 
Both datasets are gender-balanced.
Ranges denote the minimum and maximum number of utterances per speaker.}
\begin{tabular}{lcccc}
\toprule
Corpus & \#Spk (M/F) & Train & Valid & Test \\
\midrule
LibriTTS & 8 (4/4) &
$[198, 286]$ &
$[20, 29]$ &
$[20, 29]$ \\
YoBind & 6 (3/3) &
$[133, 375]$ &
$[13, 37]$ &
$[13, 37]$ \\
\bottomrule
\end{tabular}
\label{datasets_info}
\end{table}

\begin{table}[t]
\centering
\footnotesize
\caption{SECS scores of synthetic speech used as \textit{Synth 90\%}.}
\label{tab:dataset_analysis}
\begin{tabular}{l c c}
\toprule
Corpus & Fish-Speech & CosyVoice 2 \\
\midrule
LibriTTS & 0.763 & 0.794 \\
YoBind   & 0.756   & 0.788   \\
\bottomrule
\end{tabular}
\end{table}


\begin{table*}[t]
\centering
\footnotesize
\caption{Objective evaluation results on LibriTTS and YoBind datasets.
(+Extra Synth) indicates extra synthetic augmentation.
Arrows indicate whether higher ($\uparrow$) or lower ($\downarrow$) values are better.}
\label{tab:objective_results}
\setlength{\tabcolsep}{4pt}
\renewcommand{\arraystretch}{1.2}
\resizebox{\textwidth}{!}{%
\begin{tabular}{l|l|c|c|ccc|ccc}
\hline
\multirow{2}{*}{\textbf{Source ZS-TTS}} 
& \multirow{2}{*}{\textbf{Training Data}} 
& \multirow{2}{*}{\textbf{DC}} 
& \multirow{2}{*}{\textbf{OS}} 
& \multicolumn{3}{c|}{\textbf{LibriTTS}}
& \multicolumn{3}{c}{\textbf{YoBind}} \\
\cline{5-10}
& & & 
& SECS $\uparrow$ & CER $\downarrow$ & WER $\downarrow$
& SECS $\uparrow$ & CER $\downarrow$ & WER $\downarrow$ \\
\hline

\multirow{2}{*}{Real-Only Training}
& \it{Real 10\%} 
& -- & -- 
& 0.818 & 5.932 & 12.520 
& 0.813 & 5.063 & 11.035 \\

& \it{Real 100\%} 
& -- & -- 
& 0.832 & 6.539 & 13.645
& 0.840 & 4.743 & 10.424 \\
\hline

\multirow{5}{*}{Fish-Speech~\cite{liao2024fish}}

& \it{Real 10\% + Synth 90\%} 
& -- & -- 
& 0.765 & \textbf{4.738} & \textbf{10.348} 
& 0.764 & \textbf{4.081} & \textbf{9.358} \\

& \it{Real 10\% + Synth 90\%} 
& -- & \checkmark 
& 0.770 & 4.802 & 10.470 
& 0.770 & 4.085 & 9.542 \\

& \it{Real 10\% + Synth 90\%} 
& \checkmark & -- 
& 0.807 & 4.962 & 10.879
& 0.792 & 4.471 & 9.965 \\

& \it{Real 10\% + Synth 90\%} 
& \checkmark & \checkmark 
& \textbf{0.815} & 4.765 & 10.563
& \textbf{0.799} & 4.174 & 9.553 \\
\cline{2-10}

& \it{Real 10\% + Synth 90\% (+Extra Synth)} 
& \checkmark & \checkmark 
& 0.812 & 4.123 & 9.467
& 0.794 & 3.912 & 8.951 \\
\hline

\multirow{5}{*}{CosyVoice 2~\cite{du2024cosyvoice}}

& \it{Real 10\% + Synth 90\%} 
& -- & -- 
& 0.789 & \textbf{4.857} & \textbf{10.786}
& 0.774 & 4.045 & 9.065 \\

& \it{Real 10\% + Synth 90\%} 
& -- & \checkmark 
& 0.793 & 4.933 & 10.826
& 0.782 & 4.044 & 9.377 \\

& \it{Real 10\% + Synth 90\%} 
& \checkmark & -- 
& 0.808 & 5.225 & 11.345
& 0.795 & 4.279 & 9.531 \\

& \it{Real 10\% + Synth 90\%} 
& \checkmark & \checkmark 
& \textbf{0.815} & 4.943 & 10.907
& \textbf{0.804} & \textbf{4.023} & \textbf{9.006} \\
\cline{2-10}

& \it{Real 10\% + Synth 90\% (+Extra Synth)} 
& \checkmark & \checkmark 
& 0.814 & 4.560 & 10.194
& 0.805 & 3.684 & 8.364 \\
\hline

\end{tabular}
}
\end{table*}

\section{Experiments}

\subsection{Experimental Setup}
\begin{description}
  \item[\textbf{Models.}] 
We employ two open-source ZS-TTS systems with different architectures as source models to examine the generality of the proposed approach: Fish-Speech (FS)~\cite{liao2024fish} and CosyVoice 2 (CV2)~\cite{du2024cosyvoice}. 
As the target model, we adopt the VITS~\cite{kim2021conditional}, which is widely used for lightweight and efficient speech synthesis. For DC, we leverage the speaker embedding matrix of the original multi-speaker VITS implementation, with the hidden size reduced from 256 to 64, and analyze the impact of this design choice on speaker similarity in Section 4.3.

  \item[\textbf{Datasets.}] 
We use the VCTK corpus~\cite{yamagishi2019cstr} to pretrain the target TTS model, following the standard multi-speaker split adopted in the original VITS setup. 
For fine-tuning, we select eight speakers from the combined train-clean-100 and train-clean-360 subsets of LibriTTS corpus~\cite{zen2019libritts}, after removing utterances shorter than three words or containing special symbols and non-verbal tokens.
To further evaluate generalization to speakers and recording conditions not seen during ZS-TTS training, we additionally use an in-house dataset, referred to as YoBind, comprising six speakers recorded for voice assistant TTS applications.

For each target speaker, the available utterances are split into training, validation, and test sets with a ratio of 10:1:1.
To simulate a low-resource personalization scenario, only 10\% of the training utterances are randomly sampled and retained as real target-speaker recordings (\textit{Real 10\%}), while the remaining 90\% are synthesized using source ZS-TTS models (\textit{Synth 90\%}).
For zero-shot synthesis, the longest utterance among the \textit{Real 10\%} recordings that is permitted by each source model is selected as the reference prompt to maximize speaker-style coverage, based on the prior evidence that longer prompts enhance speaker similarity~\cite{giraldo2026zero}. 
In addition to this low-resource setting, we also consider a full-data fine-tuning configuration (\textit{Real 100\%}), where the target TTS model is adapted using all available real training utterances of each speaker without synthetic augmentation.

All audio samples are resampled to a sampling rate of 22 kHz and trimmed using Silero VAD~\footnote{https://github.com/snakers4/silero-vad} with a 0.2 second margin at both ends for consistency across datasets. Dataset statistics are summarized in Table~\ref{datasets_info}, and the speaker similarity characteristics of ZS-TTS synthetic speech are reported in Table~\ref{tab:dataset_analysis}.

  \item[\textbf{Training Details.}] 
For pre-training the target TTS model, we follow the training setup described in~\cite{hong2024leveraging}.
Specifically, we employ the AdamW optimizer ($\beta_1 = 0.8$, $\beta_2 = 0.99$) with an initial learning rate of $2\times10^{-4}$ decayed by a factor of $0.99118$, and pre-train the model for 400 epochs using four NVIDIA A100 GPUs.

During fine-tuning, all hyperparameters are kept identical to the pre-training setup except for the learning rate, batch size, and the number of training epochs.
The learning rate and batch size are set to $1\times10^{-5}$ and 32, respectively, and fine-tuning is performed for 600 epochs to ensure stable convergence.
OS is applied by repeating each real target-speaker utterance three times.
All fine-tuning experiments are conducted on a single A100 GPU.

  \item[\textbf{Evaluation Metrics.}] 
We evaluate synthesized speech using three objective metrics: speaker embedding cosine similarity (SECS),
character error rate (CER), and word error rate (WER).
SECS is computed as the cosine similarity between speaker embeddings extracted by the ECAPA-TDNN model~\cite{desplanques2020ecapa} trained on the VoxCeleb~\cite{nagrani2017voxceleb} and VoxCeleb2~\cite{chung2018voxceleb2} corpora, and CER and WER are measured using the Whisper medium model~\cite{radford2023robust}.
To ensure stable and reliable evaluation, we additionally include 100 randomly selected texts from the test-clean subset of LibriSpeech corpus~\cite{panayotov2015librispeech} for ASR-based metric computation.
All reported objective evaluation results are obtained by averaging over three independent runs with different random seeds for each target speaker.

We further conduct subjective evaluation with 18 listeners on randomly selected samples, using mean opinion score (MOS) to assess speech naturalness (reported for FS as a representative ZS-TTS source) and an ABX-style preference test to directly compare speaker similarity with respect to reference recordings.
In the ABX test, listeners selected the sample closer to the reference speaker, and preference percentages were computed accordingly.

\end{description}

\begin{table}[t]
\centering
\caption{Subjective evaluation results on LibriTTS and YoBind datasets.
MOS scores (95\% CI) measure perceived naturalness.
ABX values indicate preference (\%) for the proposed system over the baseline
(\textit{Real 10\% + Synth 90\%} without DC or OS). 
All ABX results are statistically significant under a two-sided binomial test ($p < 0.05$).}
\label{tab:subjective_results}

\setlength{\tabcolsep}{4pt}
\renewcommand{\arraystretch}{1.1}
\footnotesize

\resizebox{\columnwidth}{!}{%
\begin{tabular}{l|l|c|c|c|c}
\hline
\textbf{ZS-TTS} & \textbf{Training Data} & \textbf{DC} & \textbf{OS} 
& \textbf{LibriTTS} & \textbf{YoBind} \\
\hline

\multicolumn{6}{c}{\textbf{MOS (Naturalness)}} \\
\hline

\multicolumn{2}{c|}{Ground Truth (Recordings)} & -- & -- 
& 4.31 $\pm$ 0.28 
& 4.31 $\pm$ 0.30 \\
\hline

\multirow{2}{*}{Real-Only}
& \it{Real 10\%} & -- & -- 
& 3.58 $\pm$ 0.42 
& 2.85 $\pm$ 0.47 \\

& \it{Real 100\%} & -- & -- 
& 3.67 $\pm$ 0.34 
& 3.50 $\pm$ 0.40 \\
\hline

\multirow{2}{*}{FS}
& \it{Real 10\% + Synth 90\%} & -- & -- 
& 3.86 $\pm$ 0.37 
& 3.50 $\pm$ 0.35 \\

& \it{Real 10\% + Synth 90\%} & \checkmark & \checkmark 
& \textbf{3.92 $\pm$ 0.27} 
& \textbf{3.58 $\pm$ 0.38} \\
\hline

\multicolumn{6}{c}{\textbf{ABX Preference over Baseline (\%)}} \\
\hline

FS  
& \it{Real 10\% + Synth 90\%} & \checkmark & \checkmark 
& \textbf{70.8} 
& \textbf{66.7} \\

CV2 
& \it{Real 10\% + Synth 90\%} & \checkmark & \checkmark 
& \textbf{61.8} 
& \textbf{60.2} \\
\hline

\end{tabular}
}
\end{table}

\subsection{Experimental Results}

Table~\ref{tab:objective_results} shows objective evaluation results on LibriTTS and YoBind datasets with respect to the different training configurations.
Across both datasets, fine-tuning with the full set of real data (\textit{Real 100\%}) consistently yields higher speaker embedding similarity than the low-resource setting (\textit{Real 10\%}).
However, this trend is not consistently observed for ASR-based intelligibility, particularly on LibriTTS, which contains expressive audiobook-style speech with diverse prosodic patterns.
Naive mixing of zero-shot synthesized speech (\textit{Real 10\% + Synth 90\%}) substantially improves ASR-based intelligibility across both datasets, despite using the same training transcripts as the \textit{Real 100\%} setting.
This finding is consistent with prior observations that TTS models trained on synthetic data can achieve lower WER scores, potentially due in part to lower background noise and reduced variability of synthesized speech~\cite{zhou2025training,minixhofer2023evaluating}.
However, this intelligibility improvement is accompanied by a degradation in SECS, revealing a clear trade-off between intelligibility and speaker similarity.

Applying DC effectively restores speaker similarity degraded by naive synthetic augmentation, while largely preserving the intelligibility gains from synthetic data, at the cost of a slight increase in CER and WER.
OS complements DC by emphasizing scarce real target-speaker recordings and yields additional gains in speaker similarity while partially recovering intelligibility; on YoBind with CV2, it surpasses naive mixing across all metrics.
These trends are consistent across both ZS-TTS sources, suggesting that the proposed method is not tied to a specific generator.
By contrast, OS alone shows limited and unstable improvements, implying that emphasizing real samples is more reliable when synthetic-domain bias is first mitigated by DC.

To further assess realistic augmentation scenarios, we additionally synthesize extra speech using 800 randomly sampled transcripts from the VCTK corpus.
To suppress potential hallucination artifacts such as omissions and mispronunciations~\cite{zhang2025advancing} in ZS-TTS outputs, we filter synthesized speech using Whisper medium and retain only those with WER below 5\%.
Under this synthetic data scaling setting, speaker embedding similarity slightly decreases compared to the non-scaled setting, while CER and WER improve substantially.
Overall, these results suggest that the proposed approach remains effective under increased synthetic data augmentation in practical personalization scenarios.

Table~\ref{tab:subjective_results} presents subjective evaluation results. 
MOS results indicate that the proposed DC and OS strategies do not degrade overall speech quality, yielding naturalness scores comparable to those obtained with \textit{Real 100\%} fine-tuning and naive synthetic augmentation. 
In contrast, the ABX preference test shows a clear advantage of the proposed method in terms of speaker similarity.
Listeners consistently prefer speech synthesized with DC and OS over the baseline configuration (\textit{Real 10\% + Synth 90\%} without DC or OS), demonstrating that the proposed approach effectively preserves target-speaker similarity without compromising speech quality.

\begin{figure}[t]
  \centering
  \begin{subfigure}[t]{0.4\linewidth}
    \centering
    \includegraphics[width=\linewidth]{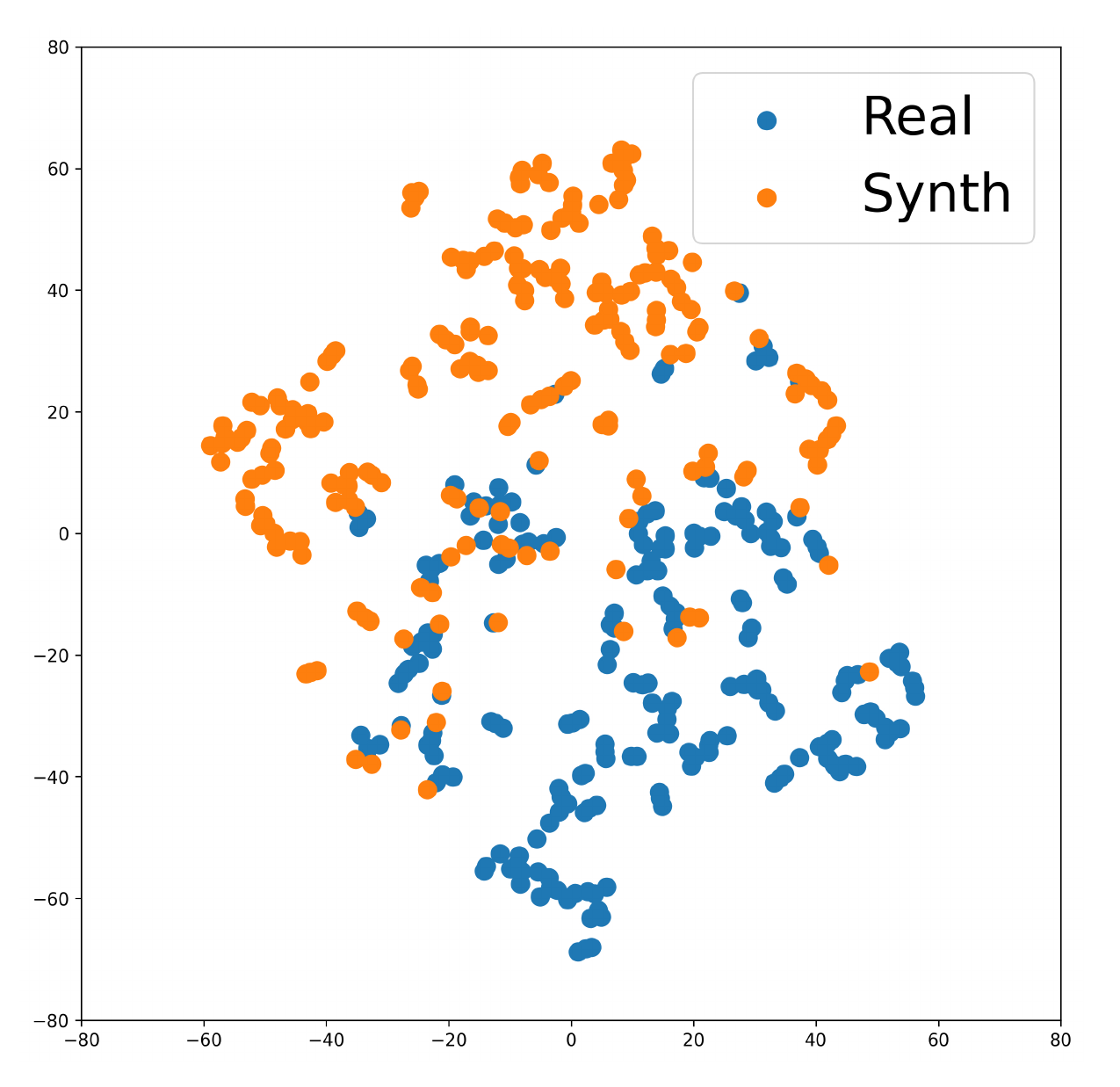}
    \caption{Speaker-matched}
    \label{fig:tsne_matched}
  \end{subfigure}
  \begin{subfigure}[t]{0.4\linewidth}
    \centering
    \includegraphics[width=\linewidth]{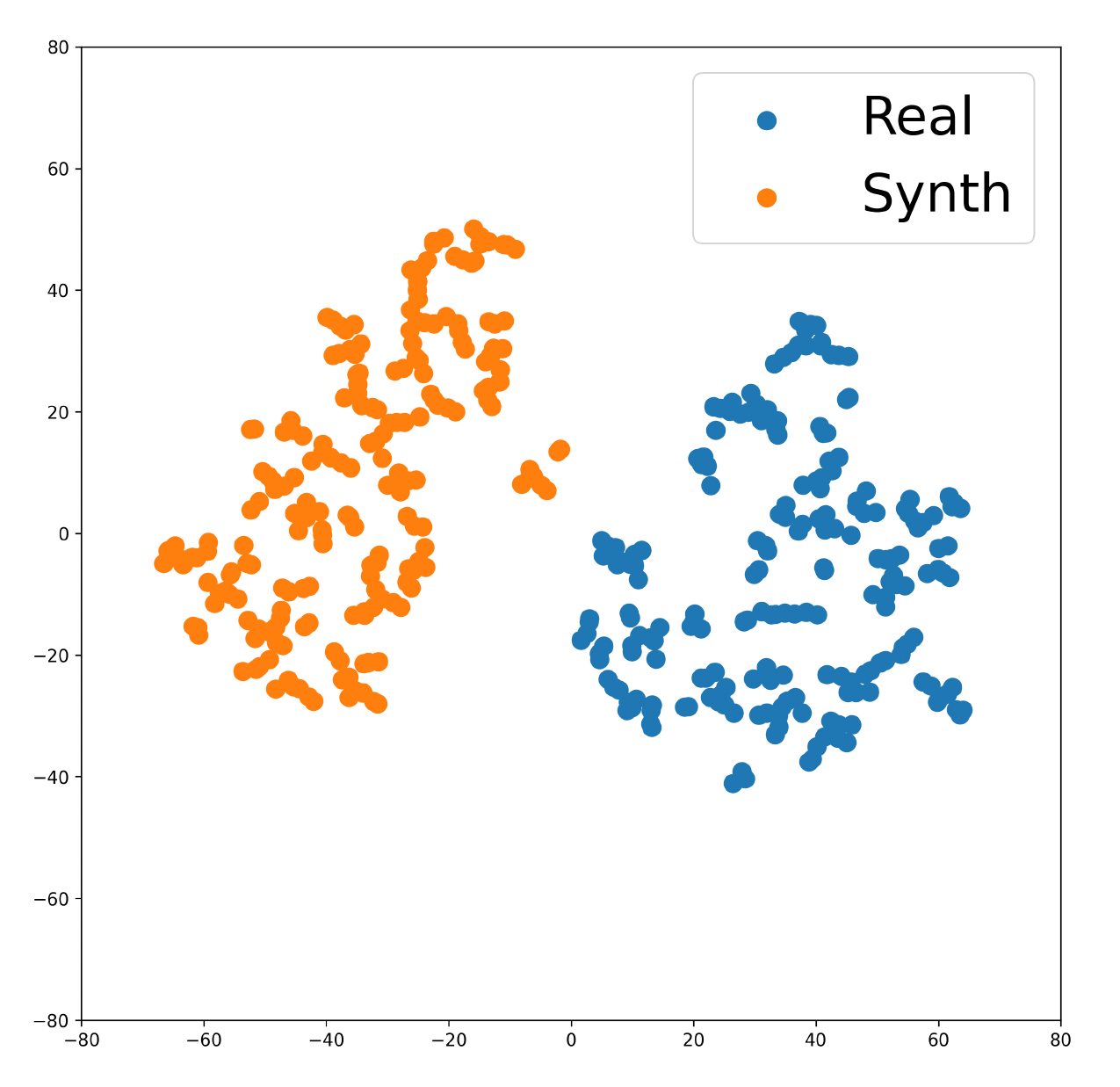}
    \caption{Speaker-mismatched}
    \label{fig:tsne_mismatched}
  \end{subfigure}
  \caption{t-SNE visualization of latent representations for a target speaker under different synthetic augmentation settings.}
  \vspace{-0.7em}
  \label{fig:tsne_comparison}
\end{figure}

\begin{table}[t]
\centering
\footnotesize
\caption{Effect of domain embedding size on speaker similarity and intelligibility
evaluated on LibriTTS.
Synthetic data are generated using Fish-Speech.
All models are trained with \textit{Real 10\% + Synth 90\%} data using
DC without OS.}
\label{tab:gin_channel_ablation}
\setlength{\tabcolsep}{6pt}
\renewcommand{\arraystretch}{1.2}
\begin{tabular}{c|ccc}
\hline
\textbf{Emb. Size}
& \textbf{SECS} $\uparrow$
& \textbf{CER} $\downarrow$
& \textbf{WER} $\downarrow$ \\
\hline
16
& \textbf{0.807}
& 5.118
& 11.229 \\
64
& \textbf{0.807}
& \textbf{4.962}
& \textbf{10.879}  \\
256
& 0.800
& 5.036
& 11.033 \\
\hline
\end{tabular}
\end{table}

\subsection{Analysis of Domain Conditioning}

Figure~\ref{fig:tsne_comparison}(a) presents the t-SNE visualization~\cite{maaten2008visualizing} of utterance-level latent representations, obtained by mean-pooling frame-level features extracted after the domain-conditioned flow module in VITS.
Real and synthetic samples exhibit partial overlap with a moderate domain shift rather than strict separation. 
This observation is consistent with the design of our domain-conditioned training, in which domain information is introduced as a controllable conditioning signal rather than being explicitly separated in the latent space.
In contrast, the speaker-mismatched augmentation setting in Figure~\ref{fig:tsne_comparison}(b), where synthetic speech is generated from a different speaker of the same gender, shows a more pronounced separation, which we analyze in detail in the following subsection.

Table~\ref{tab:gin_channel_ablation} examines the impact of domain embedding size on speaker similarity and intelligibility.
To isolate the effect of domain conditioning, experiments are conducted under DC without OS.
Performance does not improve monotonically as the embedding size increases.
A small embedding size (16) yields comparable speaker similarity but degraded intelligibility, suggesting limited capacity for effective conditioning.
Conversely, a large embedding (256) slightly reduces speaker similarity without providing further gains in ASR-based metrics.
A moderate embedding size (64) achieves the best overall trade-off, maintaining speaker similarity while substantially improving CER and WER.
These results indicate that appropriate embedding capacity is important for stable adaptation in domain-conditioned training.

\begin{table}[t]
\centering
\caption{Ablation on speaker-matched vs. speaker-mismatched synthetic data.
Synthetic augmentation uses the same setup (\textit{Real 10\% + Synth 90\%}) with DC and without OS.}
\label{tab:speaker_match}
\footnotesize
\setlength{\tabcolsep}{6pt}
\renewcommand{\arraystretch}{1.15}
\begin{tabular}{l|ccc}
\hline
\textbf{Training Data} 
& \textbf{SECS} $\uparrow$ 
& \textbf{CER} $\downarrow$
& \textbf{WER} $\downarrow$ \\
\hline

\textit{Real 10\%} (baseline)
& 0.818 
& 5.932
& 12.520 \\
\hline

Speaker-matched
& \textbf{0.807} 
& \textbf{4.962}
& \textbf{10.879} \\

Speaker-mismatched
& 0.792
& 5.817
& 12.354 \\

\hline
\end{tabular}

\end{table}

\subsection{Speaker Consistency in Synthetic Data Augmentation}

To examine whether the gains from synthetic augmentation with domain-conditioned training depend on speaker consistency rather than simply on reduced acoustic variability in ZS-TTS speech, we compare speaker-matched and speaker-mismatched synthetic data under identical training conditions.
Using Fish-Speech on LibriTTS, models are trained under the same setup (\textit{Real 10\% + Synth 90\%}) with DC and without OS, pairing the real recordings of each target speaker with either speaker-matched synthetic speech or synthetic speech from a different speaker of the same gender.
As shown in Table~\ref{tab:speaker_match}, speaker-mismatched augmentation yields only minor CER and WER improvements over the real-only baseline, but exhibits substantially lower speaker similarity than the speaker-matched setting.
This trend is consistent with the latent representations in Figure~\ref{fig:tsne_comparison}(b), where speaker-mismatched synthetic data shows stronger separation from real samples. 
Such separation suggests a larger domain gap between real and synthetic samples, which may hinder the transfer of useful linguistic information during training.
This observation highlights the importance of speaker-consistent synthetic augmentation for both speaker similarity and intelligibility.

\section{Conclusion}

We investigate the use of ZS-TTS as a data augmentation source for low-resource personalized speech synthesis and show that naively incorporating synthetic speech often degrades speaker similarity during fine-tuning.
To address this challenge, we propose ZeSTA, a simple domain-conditioned training framework with real-data oversampling that mitigates this issue without modifying the base TTS architecture. 
Through objective and subjective evaluations, we confirm that ZeSTA improves speaker similarity while retaining intelligibility gains, providing a practical strategy for controlled synthetic data integration in low-resource speaker adaptation.
Future work will explore extending ZeSTA to diverse TTS architectures and investigating architecture-specific conditioning strategies.

\section{Acknowledgments}

This research was supported by Culture, Sports and Tourism R\&D Program through the Korea Creative Content Agency grant funded by the Ministry of Culture, Sports and Tourism in 2026 (Project Name: Development of AI-based personalized cultural and arts learning services using smart device, Project Number: RS-2026-25524629, Contribution Rate: 50\%). 
This research was supported by the Startup Growth Technology Development Program (Deep-Tech TIPS) funded by the Ministry of SMEs and Startups in 2026 (Project Number: RS-2026-25542293,  Contribution Rate: 50\%).
The authors would like to thank the 18 listeners who participated in the subjective evaluation.

\section{Use of Generative AI Disclosure}

During the preparation of this manuscript, the authors used OpenAI’s ChatGPT for editing and language polishing, and the Cursor IDE for implementation-level assistance such as code refinement and debugging. These tools were used solely to improve readability, grammar, expression, and code clarity, and were not involved in the design of the proposed methods, experimental protocols, or scientific interpretations. All content, code, and conclusions were reviewed, revised, and approved by the authors, who take full responsibility for the manuscript and its submission.

\bibliographystyle{IEEEtran}
\bibliography{refs}

\end{document}